\documentclass[%
 preprint, 
superscriptaddress,
reprint,
 amsmath,amssymb,
 aps, physrev,
pra,
onecolumn,
]{revtex4-2}

\usepackage{graphicx,bm,booktabs}
\usepackage{placeins}
\usepackage[colorlinks=true,citecolor=blue,linkcolor=blue,urlcolor=blue]{hyperref}
\usepackage{cleveref}
\usepackage{xcolor}
\newcommand{\um}{\ensuremath{\,\mathrm{\mu m}}}
\newcommand{\nm}{\ensuremath{\,\mathrm{nm}}}
\newcommand{\mm}{\ensuremath{\,\mathrm{mm}}}
\newcommand{\manuscriptfigurewidth}{133.35mm}

\begin{document}
\title{Spatial–Spectral Trade-offs in Metasurface-Based Snapshot Hyperspectral Imaging}
\author{Liam Fitzpatrick}
\affiliation{%
    Department of Physics and McGill Quantum Centre, McGill University, 3600 rue University, Montréal, Québec H3A 2T8, Canada.
}%
\author{Sean Molesky}
\email{sean.molesky@polymtl.ca}
\affiliation{Department of Engineering Physics, Polytechnique Montréal, Montréal, Québec H3T 1J4, Canada.}

\author{Kai Wang}
\email{k.wang@mcgill.ca}
\affiliation{%
    Department of Physics and McGill Quantum Centre, McGill University, 3600 rue University, Montréal, Québec H3A 2T8, Canada.
}%
\date{September 23, 2026}

\begin{abstract}
    Snapshot hyperspectral imaging maps a spatial--spectral datacube onto a two-dimensional detector, with recoverability depending on the datacube sampling regime, the scene prior, and the optical encoding. We study this co-dependence with algorithmic recovery and optical design optimization in a differentiable metasurface model by varying the spatial--spectral ratio and comparing average-coherence, mutual-information, and end-to-end objectives. For compressed imaging with three bands, optimization of single-parameter dielectric pillars produces lens-like designs with overlapping wavelength-dependent point-spread functions across the objectives tested. End-to-end reconstruction can mask this collapse on restricted training distributions, showing that reconstruction fidelity depends jointly on the optical encoding and scene prior. For simple nano-pillar meta-atom geometries, the optimized singular-value spectrum deteriorates as the datacube becomes more spectrally dominated, whereas a relaxed model with independent complex transmission at each wavelength reduces this dependence. These results identify restricted wavelength-dependent transmission control in the pillar library as an important contributor to the observed spectral collapse. This motivates an illustrative local super-pixel spectrometer model in which spectral encoding is studied separately from global image formation under a plane-wave illumination approximation.
\end{abstract}

\maketitle

\section{Introduction}
Computational imaging seeks to extend the capabilities of conventional optical systems by co-designing physical measurement front-ends with computational reconstruction back-ends~\cite{Mait2018_computational_imaging,Sitzmann2018_e2e}. In compressive regimes, the number of detected quantities is smaller than the number of unknowns, so accurate recovery requires measurements matched to prior structure in the signal, such as sparsity~\cite{Donoho2006_cs,Candes2006_cs,Duarte2008_single_pixel}, model-based or structured sparsity~\cite{Baraniuk2010_model_based_cs}, and learned data priors~\cite{Bora2017_generative_cs,Ulyanov2018_deep_image_prior}. The optics both form the image, and encode information into a usable form.

Hyperspectral imaging (HSI) recovers a spatial image over many wavelength bands, producing a three-dimensional spatial--spectral datacube in which each pixel contains a spectrum. The resulting measurements support a diverse range of applications in environmental monitoring and agriculture~\cite{Bioucas2013_remote,Gowen2007_food_hsi_review,Liu2015_agri_food_hsi_review}, biosensing~\cite{Rao2022_filter_antispoofing}, and medical imaging~\cite{Lu2014_medical_hsi_review,Xiong2022_brain}.
At the same time, the additional spectral dimension creates a sampling problem: a three-dimensional datacube must be mapped onto a two-dimensional detector while preserving the spatial and spectral information relevant to the task.

Many conventional HSI systems construct the datacube by sequentially scanning a spatial or spectral coordinate~\cite{Hagen2013_hsi_rev}. Such scanning limits temporal resolution and can produce motion artifacts in dynamic scenes, while the dispersive and relay optics required can impose size and alignment costs. In contrast, \textit{snapshot hyperspectral imaging} acquires the datacube within a single detector integration period, using approaches such as coded-aperture snapshot spectral imaging and related compressive spectral cameras~\cite{Gehm2007_dual_disperser_cassi,Wagadarikar2008_single_disperser_cassi,Arce2014_cassi}. Snapshot acquisition creates a spatial-spectral sampling trade-off: in the compressed regime considered here, a datacube with $N$ unknown samples is mapped to $M<N$ detector values and must therefore exploit prior structure.

Because of their potential to support miniaturization and scalable control over spatially varying optical phase, amplitude, and wavelength response~\cite{Chen2016_metasurface_review,Li2022_meta_inverse,Roques_Carmes2025_comp_metaoptic},
optical metasurfaces are viewed as an attractive platform for realizing snapshot HSI technologies.
Existing metasurface-based spectral imagers can be understood through three characteristic spatial-spectral mixing strategies.

\textit{Local spectral-response arrays} use distinct filters or microspectrometer responses at each detector pixel or small group of pixels. Random dielectric metasurface filters have been integrated directly on detectors and combined with compressed-sensing reconstruction~\cite{Wu2022_filters_random}, reconfigurable metasurface microspectrometer arrays have enabled real-time ultraspectral imaging~\cite{Xiong2022_brain}, and trained metasurface meta-pixels have been paired with an end-to-end decoder~\cite{Makarenko2022_e2e_filters}. These systems are compact and naturally compatible with a conventional imaging lens, but spectral diversity consumes detector area and introduces a spatial-spectral resolution and crosstalk trade-off.

\textit{Spatially separated or dispersed image channels} route different wavelengths to separate sensor regions or distinct image copies. A single-layer metaoptic can be inverse-designed to demultiplex multiple channels into distinct spatial domains~\cite{Lin2022_e2e_psf}. Such channelization reduces overlap and can simplify inversion, but allocation of sensor area or aperture resources on separate channels can limit the number of bands or spatial resolution.

Finally, \textit{globally multiplexed spectral codes or wavelength-dependent point-spread functions (PSFs)} allow spectral channels to overlap in one coded sensor image and recover the datacube computationally. End-to-end metasurface and neural-reconstruction optimization has been used to exploit dispersive spectral encoding~\cite{Zhang2023_e2e_psf}. Recent work has also demonstrated metasurface spectral code masks with compressed reconstruction~\cite{Xie2026_filter}. These multiplexed systems use detector measurements efficiently, but their performance depends strongly on conditioning, calibration, noise, and the reconstruction prior. The three strategies form a continuum rather than rigid classes: a filter mosaic followed by diffraction can become a global coded measurement, while a dispersive PSF can approach explicit channel separation when wavelength responses cease to overlap.

These spatial--spectral mixing strategies are distinct from the geometry parametrization used to realize them. The mixing strategy determines how the datacube's spatial and spectral information is organized across the sensor, while the geometry parametrization maps design variables to wavelength-dependent optical responses. Rich libraries and freeform inverse-designed structures can provide stronger dispersion control, but are difficult to include in large-scale differentiable optimization~\cite{Molesky2018_inverse_design_nanophotonics,LalauKeraly2013_adjoint_shape_optimization,Li2022_meta_inverse}. Simpler parametrizations, such as single-parameter silicon pillars, are computationally convenient but provide limited spectral control. The optimization and reconstruction framework then selects and decodes an operator that the chosen parametrization can reach.
While these studies illustrate substantial promise, broader design principles of metasurface-based snapshot HSI remain largely an open question~\cite{Li2022_meta_inverse,Roques_Carmes2025_comp_metaoptic}. Reported performance depends simultaneously on the datacube sampling, the concentration of the prior, the spatial-spectral mixing strategy, the metasurface parametrization, and the reconstruction algorithm. It can therefore be difficult to determine whether strong reconstruction performance arises from optical encoding or primarily from prior-driven reconstruction. A related question is how limited, physically attainable spatial and spectral control is allocated in a flat optical system~\cite{Shrestha2018_broadband_achromatic_metalens,Presutti2020_bandwidth_metalens_limits,Liang2019_metalens_tradeoffs}. In many designs, a single-layer metasurface performs image formation and wavelength discrimination simultaneously~\cite{Lin2022_e2e_psf,Zhang2023_e2e_psf}. The compatibility of these goals under practical dispersion and parametrization limits has not been systematically clarified.

Here, we develop an operating-regime framework that separates difficulty imposed by the datacube and prior from restrictions imposed by the physical metaoptic and from choices made in optimization and reconstruction. We first organize compressed HSI using the datacube shape $\chi=N_xN_y/N_\lambda$ and compression ratio $C=N/M$, while treating prior strength as a qualitative property of the signal model. We then place coherence-based, information-theoretic, and end-to-end objectives on a common footing as different ways to preserve distinguishability under different assumptions, priors, and computational costs. We use a simplified differentiable metasurface model as a proof of principle for isolating the constraints of single-parameter pillar designs. We compare this model with independent complex transmission at each wavelength to assess how restrictions on wavelength-dependent transmission contribute to the observed encoding limitations.

In the compressed and ill-conditioned regimes, we identify a recurring optimization failure mode: with the pillar parametrization studied here, optimization tends to allocate the available degrees of freedom toward spatial imaging, converging toward a spectrally collapsed camera while sacrificing spectral encoding. Strong dataset priors can mask this collapse, yielding accurate reconstructions despite spectrally ill-conditioned measurements. Reconstruction quality alone does not necessarily imply effective physical encoding. Finally, we illustrate local spectral encoding under a plane-wave illumination approximation, motivated by separating global spatial imaging from spectral sensing. Together, these results inform design choices for compressed snapshot HSI with single-layer metasurfaces.

\section{Framework}

\begin{figure}[!htbp]
    \centering
    \includegraphics[width=\manuscriptfigurewidth]{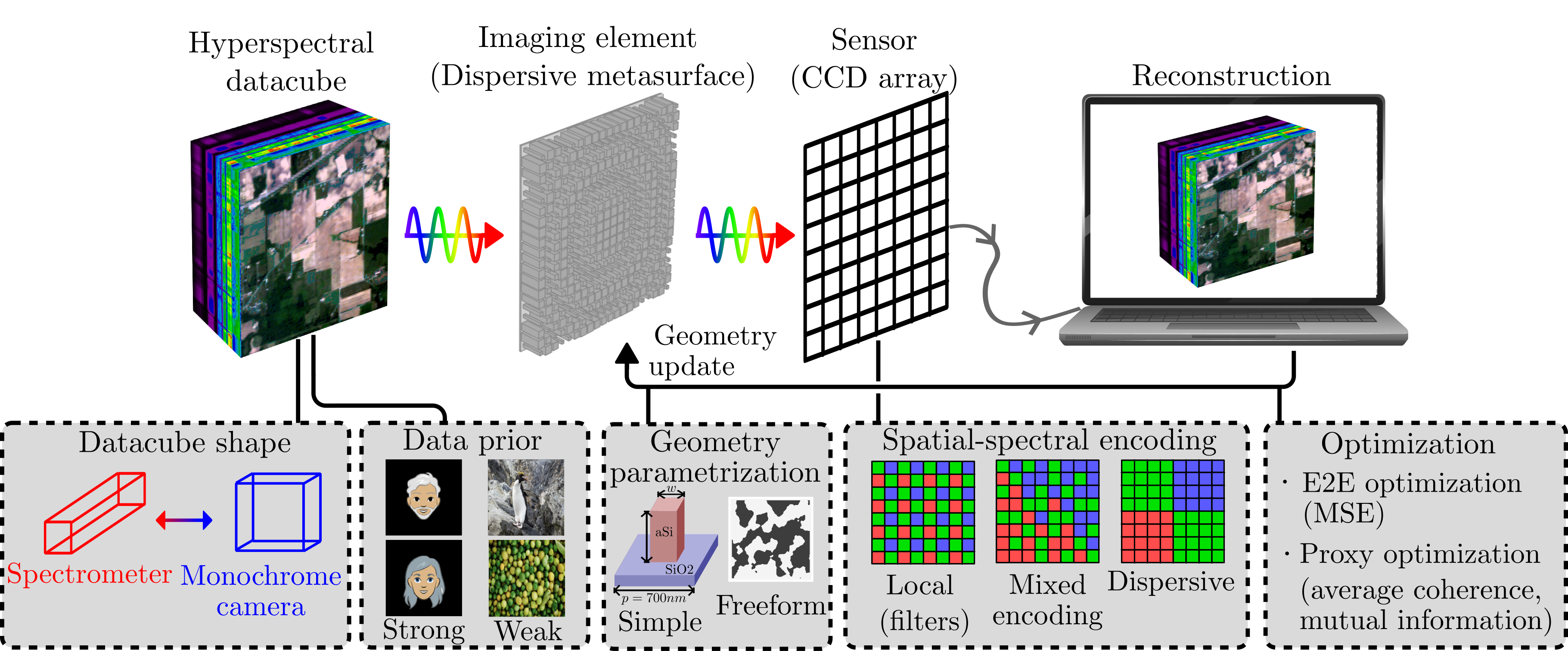}
    \caption{Overview of snapshot hyperspectral imaging with an imaging element (here, a dispersive metasurface), a sensor array, and computational reconstruction. The lower panels summarize the design axes explored in this work: datacube shape, data priors, spatial--spectral encoding strategy, and optimization framework. }
    \label{fig:problem_illustration}
\end{figure}

In this section, we establish a conceptual framework for analyzing compressed HSI, independent of the specific data representation, optical implementation, and reconstruction algorithm. We first describe the forward linear model that maps a three-dimensional scene to a two-dimensional sensor measurement. We then introduce two sampling ratios and the reconstruction prior as important factors influencing design choices. An overview of the snapshot hyperspectral-imaging problem and the design axes explored in this work is shown in Fig.~\ref{fig:problem_illustration}.

\subsection{The Compressed HSI Model}
The hyperspectral scene is represented as a three-dimensional datacube, $\mathbf{X} \in \mathbb{R}^{N_x \times N_y \times N_\lambda}$, where $N_x$ and $N_y$ are the number of spatial pixels in two dimensions, and $N_\lambda$ is the number of spectral bands. This datacube is vectorized into a single column vector $\mathbf{x} \in \mathbb{R}^{N}$, where $N = N_x \times N_y \times N_\lambda$ is the total number of points in the datacube.

A core principle of compressed sensing is that natural signals are often sparse in a suitable basis. This means that $\mathbf{x}$ can be represented as a linear combination of a few basis vectors from a dictionary or basis $\bm{\Psi} \in \mathbb{R}^{N \times K}$:
\begin{equation}
    \mathbf{x} = \bm{\Psi} \boldsymbol{\alpha},
    \label{eq:sparse_basis}
\end{equation}
where $\boldsymbol{\alpha} \in \mathbb{R}^{K}$ is a coefficient vector that is $k$-sparse, meaning it has at most $k$ non-zero elements. Here, $K$ is the number of basis vectors in $\bm{\Psi}$ and $k$ is the maximum number of nonzero coefficients in $\boldsymbol{\alpha}$.

The metasurface-based imager performs an optical transformation that maps the high-dimensional signal $\mathbf{x}$ to a lower-dimensional measurement on a conventional 2D sensor array. Under the assumption of a spatially and spectrally incoherent scene, this process can be modeled as a linear sensing operator $\bm{\Phi} \in \mathbb{R}^{M \times N}$, where $M$ is the total number of pixels on the sensor. The resulting measurement vector $\mathbf{y} \in \mathbb{R}^{M}$ is given by:
\begin{align}
    \mathbf{y} = \bm{\Phi} \mathbf{x} + \boldsymbol{\epsilon} = \mathbf{A} \boldsymbol{\alpha} + \boldsymbol{\epsilon}
    \label{eq:linear_sensing}
\end{align}
where $\boldsymbol{\epsilon}$ represents measurement noise. The operator $\bm{\Phi}$ represents the intensity response of an incoherent imaging model. The metasurface itself remains linear in complex optical field, while incoherent superposition makes the imaging problem linear in intensity. The goal of the reconstruction algorithm is to recover an estimate of the signal, $\hat{\mathbf{x}}$ (or its sparse coefficients $\hat{\boldsymbol{\alpha}}$), from the measurements $\mathbf{y}$. The matrix $\mathbf{A} = \bm{\Phi} \bm{\Psi} \in \mathbb{R}^{M \times K}$, often called the sensing matrix, is the key quantity whose properties we seek to optimize through the design of the physical sensing operator $\bm{\Phi}$. In practice, this explicit sparsity model applies to proxy objectives where $\bm{\Phi}$ is optimized with a specific prior representation $\Psi$, whereas in end-to-end optimization, the prior is learned from the training data through the selected network architecture and decoder.

The physical sampling problem is defined by $(N,M)$, where $N$ is the number of scene samples and $M$ is the number of detector measurements. When such an explicit sparse representation is used, $K$ is the number of basis vectors in $\bm{\Psi}$ and $k$ is the maximum number of nonzero coefficients in $\boldsymbol{\alpha}$. The values of $k$ and $K$ depend on the chosen representation and are not universal properties of the scene or sensor. The sparse, Gaussian, and learned-prior models are detailed in the compressed-sensing, mutual-information, and end-to-end subsections below.

\subsection{Sampling ratios and priors}
The performance of a spectral imager depends on the datacube sampling regime, the physical sensing operator, and the reconstruction prior. We use two ratios to describe the quantitative sampling regime. Prior strength is treated qualitatively because it depends on how the prior is represented in reconstruction.

\textit{Spatial--spectral ratio ($\chi = N_xN_y/N_\lambda$):} The spatial--spectral ratio compares the number of spatial degrees of freedom with the number of spectral bands. Large $\chi$ denotes a spatially dominant problem, whereas small $\chi$ denotes a spectrally dominant problem. We vary $\chi$ at fixed total problem size to separate the effects of spatial and spectral sampling.

\textit{Compression ratio ($C=N/M$):}
The compression ratio, $C=N/M$, is the ratio of scene samples to sensor measurements. When $C>1$, the unrestricted linear inverse problem is underdetermined and recovery requires additional prior information. Larger $C$ indicates a more severe dimensional bottleneck, but its value alone does not determine recoverability: rank, conditioning, noise, normalization, and alignment with the prior also matter.

\textit{Prior concentration:}
Reconstruction also depends on how narrowly the prior restricts possible signals. A narrow or concentrated prior describes a restricted, predictable signal class; a wide or diffuse prior describes a broader class and places greater demands on the physical measurement. A concentrated prior can improve reconstruction from weak measurements, but its reconstructions may not generalize beyond the training distribution. Because the prior depends on the explicit or implicit signal representation and on the reconstruction method, we use prior concentration as a qualitative descriptor.

\section{Metasurface optimization methods}

\begin{figure}[!htbp]
    \centering
    \includegraphics[width=\manuscriptfigurewidth]{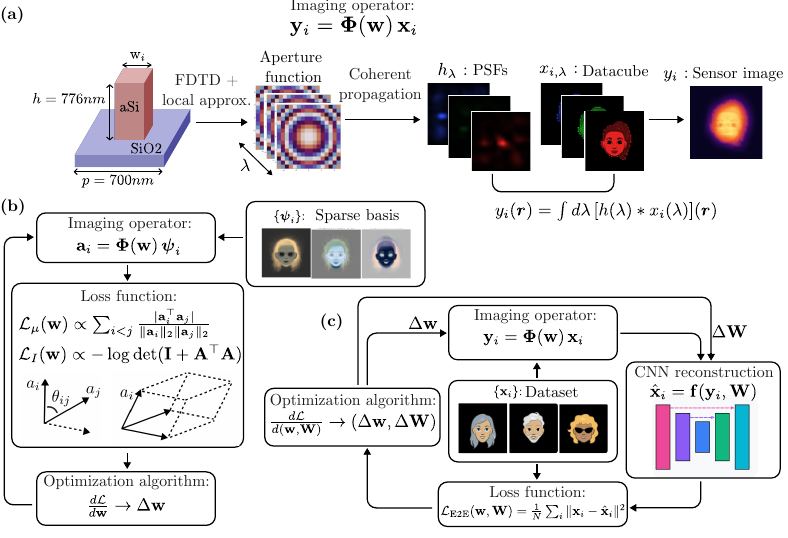}
    \caption{\textbf{(a)} Shared differentiable forward model. FDTD simulations and a local-periodic approximation map the nanopillar widths $\mathbf{w}$ to a wavelength-dependent aperture transmission. Coherent propagation with Eq. \eqref{eq:rs_propagation} produces the point-spread functions $h_\lambda$, which convolve the spectral slices $x_{i,\lambda}$. Incoherent integration over wavelength produces the detector image $y_i=\int \mathrm{d}\lambda\,[h(\lambda)*x_i(\lambda)]$. \textbf{(b)} Proxy-optimization methods. A fixed basis $\{\boldsymbol{\psi}_j\}$ is propagated through the physical operator to form sensing vectors $\mathbf{a}_j=\boldsymbol{\Phi}(\mathbf{w})\boldsymbol{\psi}_j$ and sensing matrix $\mathbf{A}=[\mathbf{a}_1,\ldots,\mathbf{a}_K]$. The cartoon-face images shown in the schematic are included for comparative illustration with the end-to-end case, the later proxy optimization experiment uses a discrete-cosine-transform (DCT) basis. The metasurface parameters $\mathbf{w}$ are optimized using the coherence and mutual-information losses shown. \textbf{(c)} End-to-end optimization. Training scenes $\{\mathbf{x}_i\}$ are passed through the shared physical operator and a CNN decoder, $\hat{\mathbf{x}}_i=\mathbf{f}(\mathbf{y}_i,\mathbf{W})$. The reconstruction loss $\mathcal{L}_{\mathrm{E2E}}$ jointly updates the optical parameters $\mathbf{w}$ and decoder weights $\mathbf{W}$.}
    \label{fig:optimization_methods}
\end{figure}

We compare three strategies for optimizing the physical sensing operator $\boldsymbol{\Phi}$: coherence-based proxies for an explicit sparse representation, mutual information under a Gaussian prior and noise model, and end-to-end training with a learned decoder. The first two methods target matrix conditioning or information content of $\mathbf{A}=\boldsymbol{\Phi}\boldsymbol{\Psi}$ for a chosen $\bm{\Psi}$. In end-to-end training, the physical operator and a nonlinear decoder are optimized jointly, with the prior determined by the dataset and represented in the learned network weights. That is, without an explicit finite dictionary $\boldsymbol{\Psi}$ or matrix $\mathbf{A}$. The resulting optical responses and reconstruction behavior are compared in the Results section. The shared imaging operator and the two optimization types are summarized in Fig.~\ref{fig:optimization_methods}.

\subsection{Imaging model}

We represent the spatially and spectrally incoherent scene intensity by
$x(\boldsymbol{r},\lambda)$. Let $t(\boldsymbol{r},\lambda)$ denote the complex field transmission of the thin metasurface for a unit-amplitude, normally incident plane wave. The field immediately after the layer is therefore $E(\boldsymbol{r},0^+,\lambda)=t(\boldsymbol{r},\lambda)$.

For the square-pillar parametrization, the complex aperture transmission is obtained from a precomputed library of finite-difference time-domain (FDTD) simulations of individual square-pillar unit cells with a $0.7\um$ pitch, a $776\nm$ pillar height, and normally incident illumination. The simulated transmitted field is normalized to a reference cell to obtain its amplitude and phase, and $t(\boldsymbol{r},\lambda)$ is then formed by interpolating this response at the side length assigned to each aperture pixel under a local-periodic approximation that neglects inter-pillar coupling.

Let $n$ be the refractive index of the homogeneous propagation medium and define $k=2\pi n/\lambda$. Using the Rayleigh--Sommerfeld (RS) impulse response and diffraction integral, the field in the sensor plane at distance $z$ is
\begin{equation}
    E(\boldsymbol{r},z, \lambda) = \int_{\mathbb{R}^2}\mathrm{d}^2\boldsymbol{r}' \, t(\boldsymbol{r}', \lambda) g_{\mathrm{RS}}(\boldsymbol{r}-\boldsymbol{r}',z, \lambda),
    \label{eq:rs_propagation}
\end{equation}
where $g_{\mathrm{RS}}(\boldsymbol{r},z,\lambda) = \frac{nz}{\mathrm{i}\lambda R^2} \left(1+\frac{\mathrm{i}}{k R}\right) \exp(\mathrm{i}k R)$, with $R=\sqrt{|\boldsymbol{r}|^2+z^2}$. The corresponding point-spread function for a fixed propagation distance $z$ is
\begin{equation}
    h(\boldsymbol{r}, \lambda) = \left|E(\boldsymbol{r},z, \lambda)\right|^2.
\end{equation}

We approximate the object as being at infinity, so the on-axis response is calculated using a plane wave incident normal to the metasurface (i.e., at $90^\circ$ to its surface). Under a paraxial, shift-invariant approximation, this PSF is applied to translated scene points and used as the convolution kernel for each spatial scene slice at a fixed wavelength. Field-angle dependence is therefore neglected. Because the scene is assumed to be mutually incoherent across both spatial positions and wavelength channels, the detector integrates intensity rather than complex field. The predicted sensor image is therefore:
\begin{equation}
    y(\boldsymbol{r}) = \int \mathrm{d}\lambda \left[h(\lambda) * x(\lambda)\right](\boldsymbol{r}) + \epsilon(\boldsymbol{r}),
    \label{eq:imaging_model}
\end{equation}
where $\ast$ denotes the spatial convolution operator, and where $\epsilon(\boldsymbol{r})$ denotes measurement noise.

We discretize the scene on a spatial grid and spectral grid, and discretize the detector on its spatial pixel grid. Let $\mathbf{x}$ stack the spatial scene samples across all wavelengths. For a given metasurface parameter vector $\mathbf{w}$, $\boldsymbol{\Phi}(\mathbf{w})$ applies the wavelength-dependent spatial convolution with $h(\boldsymbol{r},\lambda)$ to each spectral slice, sums the resulting intensities over wavelength, and samples or integrates the result on the detector grid. After discretization, this construction yields the sensing operator in Eq. \eqref{eq:linear_sensing}. In all experiments, each wavelength-dependent PSF is normalized by the power of a unit-amplitude plane wave incident on the metasurface aperture.

The model treats the metasurface as a thin, scalar, locally responding
transmission layer. It neglects polarization-dependent effects and near-field coupling between meta-atoms. Use of a single PSF for each wavelength assumes shift-invariant imaging under the paraxial approximation, with field-angle dependence and variation of the PSF across the aperture neglected.

All forward and adjoint evaluations are performed in matrix-free form. Rayleigh--Sommerfeld propagation and the wavelength-dependent PSF convolutions are implemented as padded Fast-Fourier Transform (FFT) convolutions ~\cite{Shen2006_rs_fft}. The FFTs and convolutions are evaluated with just-in-time compilation and GPU acceleration via JAX~\cite{jax2018github}, with derivatives evaluated using automatic differentiation.

\subsection{Compressed sensing: RIP and mutual/average coherence}

The ideal compressed sensing objective is that the matrix $\mathbf{A}$ acts as an isometry (distance preserving) on the set of all $k$-sparse signals. For the explicit sparse-dictionary model, $\mathbf{A}(\mathbf{w})=\boldsymbol{\Phi}(\mathbf{w})\boldsymbol{\Psi}\in\mathbb{R}^{M\times K}$. The \textit{restricted isometry property} (RIP) is a property of a matrix $\mathbf{A}$ and requires:
\begin{equation}
    (1-\delta_k(\mathbf{A}))\|\boldsymbol{\alpha}_T\|_2^2 \leq \|\mathbf{A}_T\boldsymbol{\alpha}_T\|_2^2 \leq (1+\delta_k(\mathbf{A}))\|\boldsymbol{\alpha}_T\|_2^2
\end{equation}
for every support $T\subset\{1,\ldots,K\}$ with $|T|\leq k$, where $\mathbf{A}_T$ contains the columns indexed by $T$. Here, $\delta_k(\mathbf{A})$ denotes the smallest nonnegative constant satisfying the inequalities above.

The combinatorial scaling of RIP is prohibitive for large problems. One common compressed-sensing proxy is the \textit{mutual coherence}, defined as the maximum off-diagonal normalized inner product:
\begin{equation}
    \mu_\text{max}\left(\mathbf{A}\right) = \max_{i \neq j}\, \frac{|\mathbf{a}_i^\top \mathbf{a}_j|}{\|\mathbf{a}_i\|_2 \|\mathbf{a}_j\|_2} = \max_{i \neq j}\, |\cos(\theta_{ij})|
    \label{eq:mutual_coherence}
\end{equation}
This metric quantifies the worst-case similarity/cosine between any two basis vectors after being mapped by the sensing operator. For normalized columns, it provides a bound on the RIP constant, $\delta_k\leq(k-1)\mu_\text{max}(\mathbf{A})$, and therefore connects pairwise separation to sparse-recovery guarantees. However, because the $\max$ function is non-smooth, $\mu_\text{max}$ can be difficult to optimize effectively with gradient-based methods.

To avoid this, a natural smooth alternative is simply to consider the sum of terms entering Eq.\;\eqref{eq:mutual_coherence}, which we call \textit{average coherence}. After normalizing the columns of $\mathbf{A}$, it is the mean absolute cosine similarity over unordered off-diagonal pairs:
\begin{equation}
    \mathcal{L}_\mu(\mathbf{w}) = \mu_{\text{avg}} = \frac{1}{K(K-1)/2} \sum_{i<j} \frac{|\mathbf{a}_i^\top \mathbf{a}_j|}{\|\mathbf{a}_i\|_2 \|\mathbf{a}_j\|_2}
    \label{eq:average_coherence}
\end{equation}
Minimizing average coherence encourages the columns of $\mathbf{A}$ to be, on average, as orthogonal as possible while distributing the objective across all column pairs. Unlike mutual coherence, a low average value does not by itself provide a worst-case sparse-recovery guarantee.

Equation~\eqref{eq:average_coherence} is evaluated over all unordered off-diagonal column pairs without materializing the full $K \times K$ Gram matrix $\mathbf{A}^\top\mathbf{A}$. The sensing columns are partitioned into $B$ equally sized blocks. The upper-triangular entries of each diagonal block are accumulated, and all entries of the cross-block products are accumulated. This block-wise calculation reduces memory, allowing larger problems to be run at the expense of longer runtime due to repeated calculations of the same Gram matrix entries.

\subsection{Information theory: mutual information}
An alternative approach is to frame the optimization problem from an information-theoretic perspective: we design a sensing operator $\boldsymbol{\Phi}$ to maximize the mutual information between the measurement $\mathbf{y}$ and the coefficient vector $\boldsymbol{\alpha}$ chosen from a Gaussian distribution. Assuming $\mathbf{y}=\mathbf{A}(\mathbf{w})\boldsymbol{\alpha}+\boldsymbol{\epsilon}$, with $\boldsymbol{\alpha}\sim\mathcal{N}(\boldsymbol{\mu}_\alpha,\boldsymbol{\Sigma}_\alpha)$ and $\boldsymbol{\epsilon}\sim\mathcal{N}(\mathbf{0},\boldsymbol{\Sigma}_\epsilon)$, the mutual information is:
\begin{equation}
    I(\boldsymbol{\alpha};\mathbf{y})
    = \frac{1}{2}\log\det\!\left(\mathbf{I}_K+\boldsymbol{\Sigma}_\alpha^{1/2}\mathbf{A}^\top\boldsymbol{\Sigma}_\epsilon^{-1}\mathbf{A}\boldsymbol{\Sigma}_\alpha^{1/2}\right).
    \label{eq:mutual_info}
\end{equation}
This Gaussian covariance model is the prior used for the information-theoretic objective, not an exact representation of the earlier $k$-sparse signal class.

The Fisher-information matrix (or precision matrix) for the Gaussian likelihood is $J_{\mathrm{lik}}=\mathbf{A}^{\mathsf{T}}\boldsymbol{\Sigma}_{\epsilon}^{-1}\mathbf{A}$. With the Gaussian prior, the posterior precision is $J_{\mathrm{post}}=J_{\mathrm{lik}}+\boldsymbol{\Sigma}_{\alpha}^{-1}$, so the mutual information can be interpreted as the reduction in Gaussian uncertainty volume, $I(\boldsymbol{\alpha};\mathbf{y})=\tfrac{1}{2}\log[\det\boldsymbol{\Sigma}_{\alpha}/\det\boldsymbol{\Sigma}_{\mathrm{post}}]$. The objective therefore favors sensing modes that are independent and high signal-to-noise under the assumed prior and noise model.

For the proxy objective, we assume an isotropic Gaussian prior, $\boldsymbol{\Sigma}_\alpha=\sigma_\alpha^2\mathbf{I}_K$. We also assume isotropic and uncorrelated detector noise, $\boldsymbol{\Sigma}_\epsilon=\sigma_\epsilon^2\mathbf{I}_M$. Taking the negative of the mutual information, because the quantity is maximized, gives:
\begin{equation}
    \mathcal{L}_I(\mathbf{w}) = -\log\det\!\left(\mathbf{I}_K+\frac{\sigma_\alpha^2}{\sigma_\epsilon^2}\mathbf{A}^\top \mathbf{A}\right) = -\sum_i\log\!\left(1+\frac{\sigma_\alpha^2}{\sigma_\epsilon^2}s_i^2\right),
    \label{eq:mutual_info_proxy}
\end{equation}

Here, $s_i$ denotes the singular values of $\mathbf{A}$. The singular-value form is useful because the log-determinant can be computationally expensive. The specific logarithm base is omitted because it only changes the overall objective by a constant factor.

In the implementation, the information objective is evaluated without explicitly forming $\mathbf{A}$. We apply the matrix-free sensing operator and its adjoint on a randomly chosen reduced range and compute the singular values of the resulting projected matrix, following the randomized SVD procedure~\cite{Halko2011_rsvd}. The retained singular values are then used to evaluate the mutual-information/log-determinant score. Consequently, when a truncated randomized SVD is used, the optimized quantity is a rank-reduced approximation to the full mutual-information objective rather than the exact full determinant.

\subsection{End-to-end optimization and deep learning}

Instead of optimizing a proxy metric, one can directly optimize for the final reconstruction quality. In an end-to-end approach, the metasurface's physical transformation $\boldsymbol{\Phi}$ is modeled as the first, physically constrained encoding layer, typically preceding a neural network or some learned algorithm, which is used to decode or reconstruct the scene. Unlike proxy methods, the dataset and CNN weights encode the prior instead of an explicit matrix representation of the basis $\boldsymbol{\Psi}$.

The hybrid model is trained by feeding it a dataset of known hyperspectral scenes $\{\mathbf{x}_i\}_{i=1}^N$ and corresponding measurements $\mathbf{y}_i=\boldsymbol{\Phi}(\mathbf{w}) \mathbf{x}_i+\boldsymbol{\epsilon}_i$ dependent on the metasurface parameters $\mathbf{w}$, such as widths of individual nano-pillars. The reconstruction network attempts to invert the transformation to reconstruct the original scene, $\mathbf{x}_i \approx \hat{\mathbf{x}}_i = \mathbf{f}(\mathbf{y}_i, \mathbf{W})$, where $\mathbf{W}$ represents the reconstruction-network weights. Both $\mathbf{w}$ and $\mathbf{W}$ are jointly optimized to minimize the Mean Squared Error (MSE) between the ground truth $\mathbf{x}_i$ and the network output $\hat{\mathbf{x}}_i$:
\begin{align}
    \mathcal{L}_\text{E2E}(\mathbf{w}, \mathbf{W}) = \frac{1}{N} \sum_{i=1}^N \left\| \mathbf{x}_i - \mathbf{f}\bigl(\boldsymbol{\Phi}(\mathbf{w}) \mathbf{x}_i + \boldsymbol{\epsilon}_i, \mathbf{W}\bigr) \right\|_2^2.
\end{align}

Because the physical parameters $\mathbf{w}$ are optimized using the training scenes, end-to-end co-design uses direct information about the target scene distribution when selecting the optical encoding. This method can thus theoretically achieve strong task-specific performance by jointly adapting an optical encoder and a nonlinear decoder to the training distribution, matching optically realizable channels and specific decoder architectures with the data's natural variations. This can improve task performance for a specified scene class, but makes the physical design and decoder dependent on the training distribution. Its main drawbacks are therefore the requirement of a large training dataset and the risk of poor generalization to out-of-distribution scenes.

The optical parameters and decoder weights are optimized jointly using separate Adam updates. Dataset split sizes, batch size, number of epochs, learning rates, base channel count, and sensor/metasurface grid sizes are recorded in the released code.

\subsection{Reconstruction\label{subsec:metaOpt_Recon}}

For the first two optimization methods, we need to reconstruct the signal $\mathbf{x}$ from the measurement $\mathbf{y}$. We use the following $l_1$ regularized (LASSO) minimization problem:
\begin{equation}
    \hat{\boldsymbol{\alpha}} = \arg \min_{\boldsymbol{\beta}} \|\mathbf{y} - \mathbf{A} \boldsymbol{\beta}\|_2^2 + \lambda \|\boldsymbol{\beta}\|_1
\end{equation}
where the $\ell_1$-norm is used as a tractable replacement for the $\ell_0$-norm, and where $\lambda$ is a regularization hyperparameter that controls the trade-off between measurement fidelity and compression. Larger $\lambda$ generally produces fewer and smaller coefficients, but can increase error, while smaller $\lambda$ fits the measurements more closely, but can admit noise.

The LASSO problem is solved with a matrix-free accelerated proximal-gradient method, FISTA~\cite{Beck2009_fista}. Each iteration uses one forward and one adjoint application of the sensing operator, followed by soft-thresholding in the DCT coefficient domain.

Meanwhile for the end-to-end neural network, we simply input the measurement $\mathbf{y}$ into the optimized network, $\hat{\mathbf{x}} = \mathbf{f}(\mathbf{y}; \mathbf{W})$. This allows a trained backend the possibility of fine-tuning after fabrication. Assuming sufficient stability in the co-optimized design, small fabrication or alignment errors may be corrected by conditioning the network on measured data. This flexibility is not unique to end-to-end systems: a proxy-optimized system can also be calibrated by measuring its actual $\boldsymbol{\Phi}$; however, in practice this requires a full device characterization, whereas an end-to-end system can simply be retrained with new measurements.

\section{Results}
\label{sec:results}

We use numerical optimization experiments to test how spatial and spectral information compete in a compressed metasurface imager. We first establish a recurring outcome: both proxy objectives and end-to-end optimization favor a lens-like solution with weak spectral separation. We then compare strong and weak data priors to separate the quality of the optical measurement from the quality of the learned reconstruction. Finally, we vary the datacube shape and relax the nanopillar transmission constraint to identify how restricted wavelength-dependent transmission limits the conversion of spatial oversampling into spectral encoding. The final experiment tests a task-separated architecture under the local plane-wave approximation, where a conventional lens performs global image formation and a local metasurface performs spectral encoding when single-layer dispersion is limited.

\subsection{Optical encoding and learned-prior reconstruction}

We begin with a compressed imaging problem with three near-infrared (NIR) bands. For the proxy experiments, $\bm{\Psi}$ is the complete orthonormal two-dimensional discrete-cosine-transform (DCT) basis applied independently to each of the three NIR bands. These bands are chosen as equally spaced frequencies, corresponding to wavelength samples of $1.000$, $1.130$, and $1.300~\um$.

\begin{table}[!htbp]
    \centering
    \caption{Configurations and representative PSNR values used in Fig.~\ref{fig:results_proxy_vs_e2e}. The compression ratio is $C=N/M=3$ in every row, but the spatial grid, propagation distance, data prior, and decoder differ. The PSNR values are consequently not directly comparable across the proxy and end-to-end families.}
    \label{tab:proxy_e2e_configurations}
    \small
    \setlength{\tabcolsep}{4pt}
    \resizebox{\manuscriptfigurewidth}{!}{%
        \begin{tabular}{@{}llllll@{}}
            \toprule
            Objective              & Input / Sensor & $z\, [\mu\mathrm{m}]$ & Basis $\Psi$ / dataset & Decoder & PSNR (dB) \\
            \midrule
            (a) Average coherence  & $32\times32$   & $200$                 & DCT                    & LASSO   & $9.17$    \\
            (b) Mutual information & $32\times32$   & $200$                 & DCT                    & LASSO   & $9.38$    \\
            (c) End-to-end         & $128\times128$ & $2000$                & CartoonSet             & CNN     & $36.83$   \\
            (d) End-to-end         & $128\times128$ & $2000$                & DIV2K                  & CNN     & $16.10$   \\
            \bottomrule
        \end{tabular}%
    }
\end{table}

\begin{figure}[!htbp]
    \centering
    \includegraphics[width=\manuscriptfigurewidth]{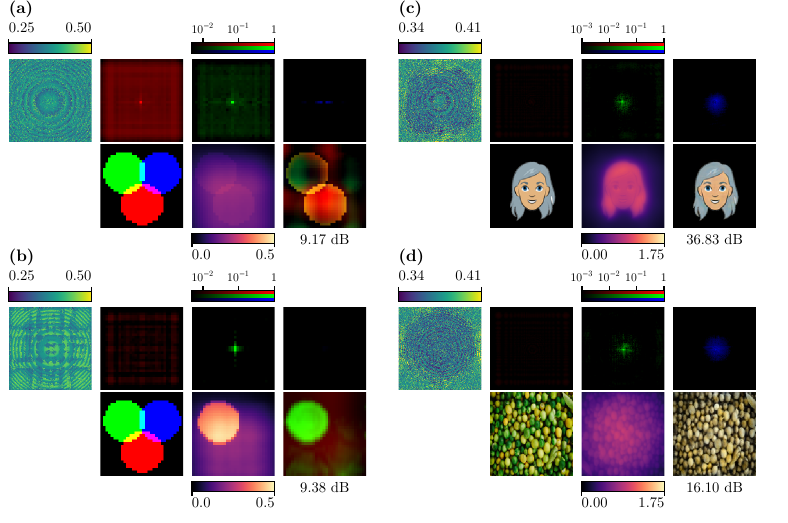}
    \caption{Proxy methods (left) and end-to-end methods (right). Each panel places optimized metasurface side lengths and three sensor-plane PSFs above the input, sensor measurement, and reconstruction. PSFs are false-color red, green, and blue at $1.300$, $1.130$, and $1.000~\mu\mathrm{m}$. A three-strip logarithmic colorbar above the green PSF gives the common scale. Results are shown for \textbf{(a)} average-coherence proxy optimization using a $32\times32$ input/sensor grid at $z=200~\mu\mathrm{m}$ with DCT/LASSO reconstruction, \textbf{(b)} mutual-information proxy optimization using the same configuration, \textbf{(c)} end-to-end optimization with a trained CNN decoder on the CartoonSet dataset using a $128\times128$ input/sensor grid at $z=2000~\mu\mathrm{m}$, and \textbf{(d)} the same end-to-end configuration trained on DIV2K. Although $C=3$ in all four cases, the differing grid sizes, propagation distances, scene distributions, and decoders make the PSNR values unsuitable for direct cross-family comparison. For both the metasurface grid and the sensor grid, $x,y\in[-56,56]~\mu\mathrm{m}$ in panels (a,b) and $x,y\in[-224,224]~\mu\mathrm{m}$ in panels (c,d); the PSFs use the same sensor-grid extents. Side-length colorbars are above each side-length image, and sensor colorbars are below each sensor measurement, with scales shared within each pair. Pillar-width display ranges are $0.25$--$0.50~\mu\mathrm{m}$ for (a,b) and $0.34$--$0.41~\mu\mathrm{m}$ for (c,d); the narrower end-to-end range reveals the low-contrast focusing pattern.}
    \label{fig:results_proxy_vs_e2e}
\end{figure}

The plotted examples use two different spatial-grid configurations. The proxy grids use $N=3\times32\times32$ and $M=32\times32$, while the end-to-end grids use $N=3\times128\times128$ and $M=128\times128$. Thus $C=N/M=3$ in both cases, but the spatial sampling and physical scale are different. These parameters were chosen due to poor computational scaling for the proxy methods at higher spatial resolutions. Table~\ref{tab:proxy_e2e_configurations} summarizes the conditions and representative PSNR values for each experiment.

The optimization results for average coherence and mutual information (MI) are shown in Fig.~\ref{fig:results_proxy_vs_e2e}(a)--(b). Both methods produce lens-like designs, but the spatial response varies across wavelengths. Average-coherence optimization focuses the $1.300$ and $1.130~\um$ channels similarly, whereas MI optimization concentrates focusing in the $1.130~\um$ channel and leaves the other two channels largely defocused. These wavelength-dependent responses do not, by themselves, establish effective spectral encoding of an extended scene, where contributions from different spatial positions also overlap.

We next ask whether the same optical outcome is specific to proxy optimization. We jointly optimize the metasurface and a CNN decoder on two three-channel image datasets with different prior strengths: CartoonSet, a collection of procedurally generated cartoon avatars~\cite{Royer2020_xgan_cartoonset}, and DIV2K, a diverse natural-image dataset~\cite{Agustsson2017_DIV2K}. We use $128\times128$ images, with 4096 training and 512 validation examples from CartoonSet and the 800-image training and 100-image validation splits of DIV2K. In the optical model, the three channels are interpreted as the three NIR bands described above rather than as visible-light channels.

For these end-to-end experiments, we use a fully convolutional reconstruction network for the three NIR bands. The scalar detector image is expanded to one channel and processed by four same-padded ReLU convolutions with channel widths $C$, $2C$, $4C$, and $8C$ and kernel sizes $7\times7$, $5\times5$, $3\times3$, and $3\times3$, respectively. A decoder-side sequence uses widths $8C$, $4C$, $2C$, and $C$ with kernel sizes $3\times3$, $3\times3$, $3\times3$, and $5\times5$. A final same-padded $7\times7$ convolution produces the three reconstructed channels. In addition, $1\times1$ projections of the first two feature maps are added to the output with coefficient $0.1$ as skip connections.

This comparison has two purposes. First, it tests whether end-to-end training changes the qualitative optical solution selected by the proxy methods. Second, it tests whether a learned prior can produce an accurate reconstruction even when the optical measurement has weak spectral discrimination. It is thus important to interpret the reconstruction results alongside the optical characterization.

The end-to-end results are shown in Fig.~\ref{fig:results_proxy_vs_e2e}(c)--(d). Both datasets produce a lens-like, weakly spectrally discriminating optical design similar to the proxy solutions, however the reconstruction results differ sharply. For the broad DIV2K distribution, the decoder recovers recognizable spatial structure but has substantial band-reconstruction error, consistent with the optical measurement lacking enough independent NIR information. For the constrained CartoonSet distribution, the decoder produces accurate in-distribution reconstructions despite the same optical collapse. This contrast is consistent with a narrow learned prior supporting reconstruction despite limited optical spectral discrimination, and motivates assessing optical encoding alongside reconstruction fidelity.

We also briefly explored an intermediate approach based on dictionary learning, where a sparse basis $\Psi$ is learned separately on the dataset and proxy methods are then used to optimize over that basis. This provides a bridge between explicit bases for proxy optimization and implicit nonlinear priors in end-to-end learning. In practice, this approach quickly became computationally prohibitive as the image size increased, since the learned dictionary had to be stored in full in memory and single matrix-vector products could not be computed as is possible for a basis such as the DCT with a defined integral transform. This scaling made it difficult to apply proxy-based optimization with learned dictionaries for useful sensor sizes, motivating end-to-end learning for these larger dataset-specific problems.

Across the proxy and end-to-end experiments, the simple pillar parametrization produces similar lens-like optical responses with weak spectral discrimination, while reconstruction quality varies with the learned prior. The next section examines how relaxing the pillar library's wavelength-dependent transmission constraints changes the optimized sensing operator.

\subsection{Impact of datacube shape on optical collapse}
\label{sec:shape_ratios}

To probe the operating regimes of the optimization framework, we analyzed the performance of the metasurface across four distinct datacube shapes ($\chi$) while maintaining a constant $N$ and $M$. We fixed the problem size to $N=256$ unknowns and used a $16 \times 16$ sensor with $M=256$ pixels, giving a fixed compression ratio $C=N/M=1$. This is not dimensionally compressed sensing, however, imperfect and overlapping PSFs make the square sensing operator poorly conditioned, making recovery ill-conditioned. We compare the performance of the metasurface across four different datacube shape ratios: $\chi = 256$ ($16 \times 16$ spatial scene, 1 spectral band), $\chi = 16$ ($8 \times 8$ spatial, 4 bands), $\chi = 1$ ($4 \times 4$ spatial, 16 bands), and $\chi = 1/16$ ($2 \times 2$ spatial, 64 bands).

Because $N$ and $M$ are fixed while the spatial grid changes, the detector oversampling factor $q=M/(N_xN_y)$ takes values $1$, $4$, $16$, and $64$ for $\chi=256$, $16$, $1$, and $1/16$, respectively, matching the spectral sampling in each case. The multiband cases all span $1.1$--$1.5\um$, whereas the single-band case samples $1.1\um$. Decreasing $\chi$ therefore simultaneously reduces spatial sampling, increases detector samples per spatial mode, and decreases $\Delta\lambda$, creating a trade-off between spatial and spectral resolution.

For every shape and propagation distance, we evaluate the mutual-information objective with the full singular-value decomposition of the resulting $256\times256$ sensing matrix. We retain the full rank because the monochromatic $\chi=256$ case can approach a nearly full-rank, lens-like spatial imaging operator. Truncating the objective would change the optimization target by rewarding only a subset of its measurement modes.

For each case, we optimized the metasurface using the mutual-information metric across a sweep of propagation distances $z$, ranging from $1\um$ to $1\mm$. We assume unit coefficient and detector-noise covariances, so the optimization objective is $-\sum_i\log(1+s_i^2)$. The square-pillar library used in the constrained experiments spans side lengths from $0.1$ to $0.65\um$ and wavelengths from $0.8$ to $1.7\um$.

\begin{figure}[!htbp]
    \centering
    \includegraphics[width=\manuscriptfigurewidth]{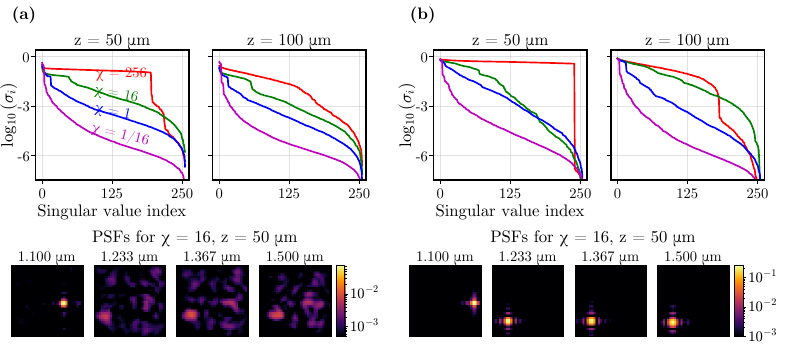}
    \caption{\textbf{(a)} Metasurface optimization results for simple square nano-pillar geometry versus \textbf{(b)} optimization results for unconstrained complex transmission values. Top: singular-value distributions for different datacube shapes $\chi$ and propagation distances $z$. Bottom: optimized PSFs for both panels at $\chi=16$ and $z=50\um$. Every PSF shows the central detector region with $x,y\in[-11.2,11.2]~\mu\mathrm{m}$.}
    \label{fig:svd_shape_sweep}
\end{figure}

Fig.~\ref{fig:svd_shape_sweep}(a) presents the singular-value spectra of the optimized sensing matrices $A$ and their PSFs for the case of $\chi=16$ with 4 spectral bands. A clear physical hierarchy emerges from the results. In the spatially-dominant regime, the singular-value spectrum remains relatively flat and better conditioned than in the spectrally dominant cases. The metasurface, composed of simple dielectric pillars, acts effectively as a lens, mapping spatial modes to distinct sensor locations with high orthogonality.

\begin{table}[!htbp]
    \centering
    \caption{Quantitative summary of the optimized shape sweep at $z=50\um$. Paired entries are square-pillar/independent-complex results. The wavelength-range column gives the sampled endpoints for each case, and $\Delta\lambda$ reports the spectral spacing on the interval. The positive information score $-\mathcal{L}_I$ follows Eq.~\eqref{eq:mutual_info_proxy}, and $G_I=(-\mathcal{L}_I)_{\mathrm{complex}}/(-\mathcal{L}_I)_{\mathrm{pillar}}$ is the complex-to-pillar score ratio.}
    \label{tab:svd_shape_sweep}
    \normalsize
    \setlength{\tabcolsep}{4pt}
    \vspace{3pt}
    \begin{tabular}{@{}cccccc@{}}
        \toprule
        $\chi$ & $N_x\!\times\!N_y\!\times\!N_\lambda$ & $\lambda$ range    & $\Delta\lambda$ & $-\mathcal{L}_I$ & $G_I$   \\
        \midrule
        $256$  & $16\!\times\!16\!\times\!1$     & $1.1\um$--$1.1\um$ & ---             & $4.574/52.998$   & $11.59$ \\
        $16$   & $8\!\times\!8\!\times\!4$       & $1.1\um$--$1.5\um$ & $133.3\nm$      & $0.372/8.827$    & $23.72$ \\
        $1$    & $4\!\times\!4\!\times\!16$      & $1.1\um$--$1.5\um$ & $26.7\nm$       & $0.284/2.873$    & $10.11$ \\
        $1/16$ & $2\!\times\!2\!\times\!64$      & $1.1\um$--$1.5\um$ & $6.35\nm$       & $0.425/1.458$    & $3.43$  \\
        \bottomrule
    \end{tabular}
\end{table}

Conversely, as the problem becomes spectrally dominant, the singular values collapse rapidly. We hypothesize this is partially because the simple square-pillar model provides limited wavelength-dependent control and therefore converts only part of the available oversampling into independent spectral directions. These results show an operating regime in which this metasurface is an efficient spatial encoder but an inefficient spectral encoder.

To test this hypothesis, we repeated the sweep with a fictitious complex-transmission model in which each design location and wavelength has an independent field transmission $t=a e^{i\varphi}$, with amplitude $a\in[0,1]$ and phase $\varphi$ modulo $2\pi$. Because the responses are chosen independently at each wavelength, this model need not satisfy the causal dispersion constraints of a realizable broadband material. It is therefore used only as a diagnostic relaxation. Fig.~\ref{fig:svd_shape_sweep}(b) shows the results of this sweep. Table~\ref{tab:svd_shape_sweep} summarizes the shape sweep and information objective at $z=50\um$.

Relaxing the pillar-constrained transmission model broadens the singular spectrum, increases $-\mathcal{L}_I$ across all four datacube shapes, and avoids overlapping PSFs indicative of spectral collapse. This shows that independent wavelength control converts more of the available detector oversampling into independent measurement directions. The gain is largest at $\chi=16$ and remains positive at $\chi=1/16$, although the absolute information score still decreases as the problem becomes more spectrally dense.

\subsection{Motivating the hyperspectral super-pixel imager}

\begin{figure}[!htbp]
    \centering
    \includegraphics[width=\manuscriptfigurewidth]{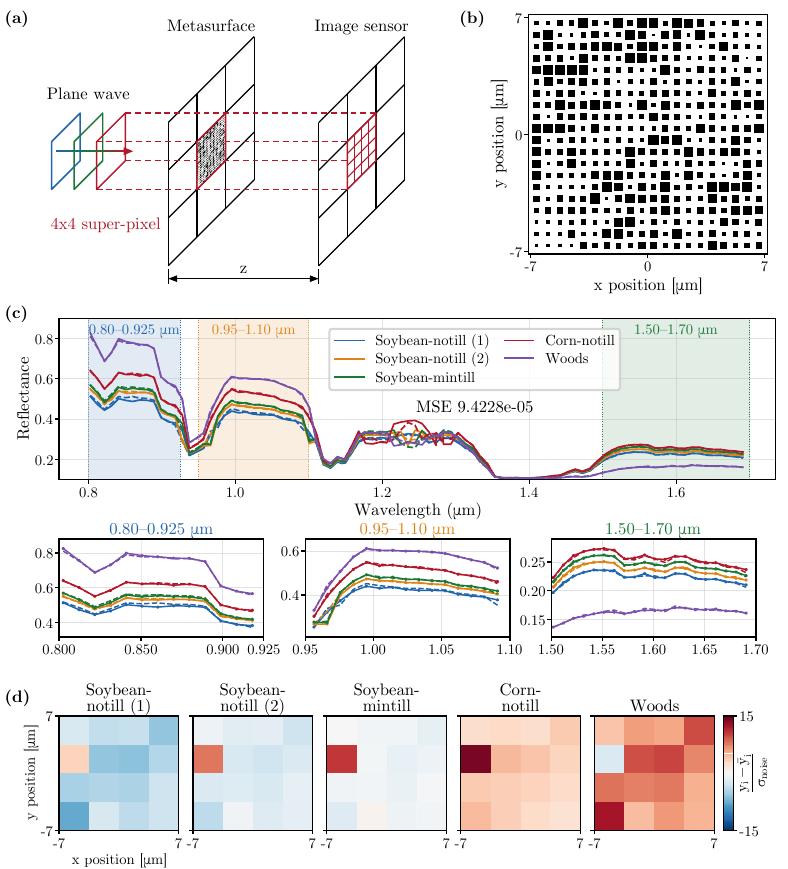}
    \caption{Superpixel spectrometer. \textbf{(a)} A spatially coherent and spectrally incoherent plane wave is encoded by a metasurface super-pixel and measured by a $4\times4$ detector patch located $z\approx10\,\mu\mathrm{m}$ downstream. \textbf{(b)} End-to-end optimized $20\times20$ square-pillar metasurface. Black-square side lengths indicate optimized pillar widths with a $0.7\,\mu\mathrm{m}$ lattice pitch. \textbf{(c)} Target spectra (solid curves) and multilayer-perceptron reconstructions (dashed curves) for five illustrative AVIRIS Indian Pines samples~\cite{Baumgardner2015_indian_pines}, highlighting 3 bands with the greatest variation in the data. The reported MSE is over the validation set, with displayed samples chosen for illustration. \textbf{(d)} Corresponding $4\times4$ detector responses, after subtracting the training-set mean response and dividing each sample by its noise standard deviation, $\sigma_{\mathrm{noise}}$. The experiment demonstrates local spectral encoding under the plane-wave approximation and does not include global image formation or inter-superpixel crosstalk.}
    \label{fig:superpixel}
\end{figure}

In this architecture, the large problem is partitioned into an array of independent microscopic problems. Each "super-pixel" on the sensor array corresponds to a single spatial location on the scene but creates a local encoding problem with a short propagation distance ($z \approx 10\um$), treating each super-pixel as an independent spectral sensor. The problem is therefore not a full hyperspectral imaging problem, but instead a local spectrometry problem: a one-dimensional input spectrum is mapped to a small $4\times4$ sensor patch and reconstructed by a learned network. The metasurface and multilayer perceptron (MLP) backend are jointly optimized end-to-end using the MSE of the reconstructed dataset.

We use the AVIRIS Indian Pines hyperspectral scene, a $145\times145$ image with 220 spectral bands, showing multiple different crop/vegetation classes in distinct spatial regions~\cite{Baumgardner2015_indian_pines}. We retain the 94 spectral samples supported by the $0.8$--$1.7~\um$ optical range and use a 95\%/5\% training/validation split for the super-pixel experiment. The flattened detector response is then processed by a fully connected MLP with hidden widths 128, 256, 384, and 384. Each hidden layer consists of a learned affine transformation followed by a ReLU activation. A final linear layer maps the 384-dimensional feature vector to the reconstructed spectrum over the full 94 wavelength channels. The optimization results and reconstruction examples are shown in Fig.~\ref{fig:superpixel}(b)--(d).

The optimized super-pixel produces wavelength-dependent detector responses, as shown in Fig.~\ref{fig:superpixel}(d). After subtracting the training-set mean response and whitening by the detector-noise scale, several measured values exceed one in magnitude. This indicates that the local optical response varies appreciably with wavelength relative to the adopted noise level, providing evidence of discrimination between vegetation classes. The MLP reconstructs spectra from these measurements using the learned scene prior. We therefore interpret the experiment as a demonstration of task-specific local spectral reconstruction, rather than as evidence that all 94 spectral channels are independently encoded by the optics.

Several approximations are made in this model. First, we assume that image formation has already been performed by a conventional lens, and at each super-pixel region, the wavefront can be approximated as a locally coherent plane wave. We also use a short propagation distance between metasurface and sensor to confine the diffracted light and minimize crosstalk between adjacent super-pixels. These assumptions define the scope of the local spectrometer model; their validity in a complete imaging system requires separate evaluation. The example therefore illustrates local spectral encoding under plane-wave illumination rather than a complete snapshot hyperspectral camera.

\section{Conclusions}

We have organized metasurface-based snapshot hyperspectral imaging in terms of the compression ratio $C$, the spatial--spectral datacube shape $\chi$, and the concentration of the reconstruction prior. These quantities separate the dimensional bottleneck imposed by the imaging problem from the wavelength-dependent transformations available to the optical design.

Across the compressed three-band proxy experiments, average-coherence and mutual-information optimization produced lens-like metasurfaces with strongly overlapping spectral point-spread functions. The optimization therefore retained dominant spatial information while providing weak independent spectral encoding. This is the recurring optical collapse identified in the underdetermined imaging regime.

The end-to-end experiments showed that reconstruction quality is not, by itself, evidence of informative spectral encoding. End-to-end training produced a similar lens-like optical design for both datasets, but a narrow dataset prior supported accurate in-distribution reconstruction whereas the broader natural-image dataset exposed substantial band-reconstruction error. Optical conditioning and wavelength discrimination should therefore be reported alongside reconstruction metrics, particularly when a learned decoder is used.

The datacube-shape sweep then showed how this behavior depends on the spatial--spectral allocation. Single-parameter dielectric pillars retained spatial modes effectively, but the singular-value spectrum deteriorated as spectral sampling became denser. Independent complex transmission at each wavelength greatly reduced this deterioration in the intermediate regimes, showing that restricted spectral response limits how effectively available detector oversampling is converted into spectral rank. These results highlight the importance of wavelength-dependent transmission control alongside the choice of optimization objective and decoder.

Motivated by these observations, we illustrated local spectral encoding with a metasurface super-pixel under plane-wave illumination, with global image formation assigned to a separate optical element. The demonstrated model is a local spectrometer rather than a complete hyperspectral camera. Extension to an imaging system requires evaluating inter-superpixel diffraction and the angular and coherence properties of the illumination supplied by the imaging optics. Within these limits, the results motivate task-separated optical architectures as promising approaches to snapshot hyperspectral imaging with physically constrained metasurface platforms.

\appendix

\begin{acknowledgments}
    We acknowledge support from the Optica Foundation and Natural Sciences and Engineering Research Council of Canada (NSERC) under RGPIN-2023-03630 and RGPIN-2023-05818.
    L.F. is partially supported by the Advanced Materials Academy, an NSERC CREATE program.
    This research was undertaken, in part, thanks to funding from the Canada Research Chairs Program (CRC-2024-00338).
     A part of the computations were performed on the digital research infrastructure provided by the Digital Research Alliance of Canada (dkx-175-ab).
    This work benefited from the research communities fostered by the Institut transdisciplinaire d'information quantique (INTRIQ) and the Regroupement québécois sur les matériaux de pointe (RQMP), both part of the Fonds de recherche
    du Qu\'ebec - Nature et technologie (FRQNT) Regroupements stratégiques program.

\end{acknowledgments}

\section*{Disclosures}
The authors declare no conflicts of interest.

During preparation of this manuscript, the authors used ChatGPT 5.6/6 in codex to debug, optimize, and better organize python simulation code for numerical experiments. The authors reviewed and tested all AI-assisted code against verifiable tests. The authors used ChatGPT 5.6/6 to assist with grammar and to organize ideas originally drafted by the authors. They were not used to generate ideas/conclusions or substantive manuscript text. No AI tool was used to create or alter research data, figures or illustrations. The authors take full responsibility for the manuscript's content, results, and conclusions.
\bibliography{references}

\end{document}